\documentclass[12pt]{article}

\usepackage[english]{babel}
\usepackage{amsmath,amsthm,amssymb,amsfonts,url}
\usepackage{mathptmx}      

\usepackage{color}

\newtheorem{definition}{Definition}

\begin{document}

\title{Reduction of balance laws in (3+1)--dimensions to autonomous
conservation laws by means of equivalence transformations}

\author{M.~Gorgone, F.~Oliveri and M.~P.~Speciale\\
\ \\
{\footnotesize MIFT Department, University of Messina}\\
{\footnotesize Viale F. Stagno d'Alcontres 31, 98166 Messina, Italy}\\
{\footnotesize mgorgone@unime.it; foliveri@unime.it; mpspeciale@unime.it}
}

\date{Published in \textit{Acta Appl. Math.} \textbf{132}, 333--345 (2014).}

\maketitle

\begin{abstract}

A class of partial differential equations (a conservation law and four balance laws),
with four independent variables and involving sixteen arbitrary continuously differentiable functions, is considered in the
framework of equivalence transformations.
These  are point transformations  of differential equations involving arbitrary elements
and live in an augmented space of independent,
dependent and additional variables representing values taken by the arbitrary elements.
Projecting the admitted symmetries into the space of independent and dependent variables, we determine some
finite transformations mapping the system of balance laws to an equivalent one with the same differential structure but involving different arbitrary
elements; in particular, the target system we want to recover is an autonomous system of conservation laws. An application to a physical problem is considered.
\end{abstract}


\maketitle
 
\section{Introduction}
\label{sec:introduction}
Physical laws are often expressed mathematically by systems of partial differential equations (PDEs)
in the form of balance laws \cite{Dafermos,Bressan},
\begin{equation}
\label{balance}
\sum_{i=1}^n\frac{\partial\mathbf{F}^i(\mathbf{u})}{\partial x_i}=\mathbf{G}(\mathbf{u}),
\end{equation}
where $\mathbf{u}\in \mathbb{R}^m$ denotes the set of unknown fields, $\mathbf{x}\in \mathbb{R}^n$
the set of independent variables,
$\mathbf{F}^i(\mathbf{u})$ the components of a flux, and $\mathbf{G}(\mathbf{u})$ the production
term; when $\mathbf{G}(\mathbf{u})\equiv\mathbf{0}$, we have a system of conservation laws.
In this paper, the first component $x_1$ of the independent variables is the time,
and the components of $\mathbf{F}^1$ are the densities of some physical quantities.
The presence of the source terms in
systems in divergence form implies additional mathematical difficulties
in solving various problems. 
For instance, from a numerical point of view, the presence of source terms may require fractional 
step splitting methods where one alternates between solving a homogeneous system of conservation laws and an ordinary differential system obtained from the system of balance laws by dropping the terms involving space derivatives. It is known \cite{LeVeque} that for some type of problems fractional step splitting methods perform quite poorly.
Systems like (\ref{balance}) fall in the more general class of nonhomogeneous
quasilinear first order systems of PDEs.
Further mathematical difficulties may arise in those problems where the coefficients involved in the differential equations depend also
on the independent variables $\mathbf{x}$, accounting for material inhomogeneities, or special geometric assumptions, or external actions.

In dealing with differential equations, Lie group theory
\cite{Ovsiannikov,Ibragimov,Olver,Ibragimov:CRC,Olver2,Baumann,Meleshko2005,BlumanCheviakovAnco}
yields general algorithmic methods either for the determination of special
(invariant) solutions \cite{OliveriSpeciale1998,OliveriSpeciale1999,OliveriSpeciale2002,OliveriSpeciale2005,MargheritiSpeciale} of initial and boundary value problems, or the derivation of conserved quantities, or the construction of relations between different differential
equations that turn out to be equivalent
\cite{BlumanCheviakovAnco,KumeiBluman,DonatoOliveri1993,DonatoOliveri1994,DonatoOliveri1995,DonatoOliveri1996,CurroOliveri2008,Oliveri2010,Oliveri2012}.

In this paper, in the context of equivalence transformations of differential equations \cite{Ovsiannikov,IbragimovTorrisiValenti,Lisle,IbragimovTorrisi1994,TorrisiTracinaproceedings1996,Meleshko1996,TorrisiTracina1998,OzerSuhubi}, we  consider a
$(3+1)$--dimensional system of first order PDEs consisting of a linear conservation law and four general balance laws involving some
arbitrary functions.
The aim is to identify classes of systems that can be mapped through an invertible point transformation to a system of autonomous conservation laws.
Recently \cite{Oliveri2012}, it has been shown that the transformation of a general nonautonomous and/or nonhomogeneous first order quasilinear system of PDEs (which every system of first order balance laws reduces to) into autonomous and homogeneous quasilinear form is possible if and only if a suitable algebra of point symmetries is admitted. The theorem proved in 
Ref.~\cite{Oliveri2012} generalizes a theorem established in Ref.~\cite{CurroOliveri2008} for $2\times 2$ quasilinear first order systems.
Both theorems may be applied when we consider a given system of PDEs and the required hypotheses are fulfilled.
If one is interested to identify the systems of balance laws (possibly nonautonomous)
that can be transformed by an invertible point transformation to an autonomous  system of conservation laws, a convenient approach consists in using equivalence transformations.
A similar approach has been used recently in \cite{OliveriSpeciale2012} for a $2\times 2$ first order quasilinear system of PDEs, and in \cite{OliveriSpeciale2013} for a system of three balance laws in three independent variables.

The plan of the paper is the following. In Section~\ref{liesymmetries}, we recall the very basic elements concerning the Lie symmetries
of differential equations; also, the main ideas about equivalence transformations of differential equations (and the way they are used) are introduced.
In Section~\ref{model}, we investigate a class of differential equations, involving four independent and five dependent variables, expressed under the form
of a linear conservation law and four nonlinear balance laws; the considered system involves sixteen arbitrary functions of the independent and dependent variables. The equivalence
transformations are determined and the finite transformations generated by the admitted generators are constructed. As a consequence, the equivalent conservation laws are
characterized. An example of physical interest (3D Euler equations for an ideal gas in a non--inertial frame and subject to gravity) is also considered in Section~\ref{applications}.

\section{Equivalence transformations}
\label{liesymmetries}
In this Section, to fix the notation and to render the paper self--contained, we briefly recall the main elements of Lie group analysis of differential
equations, and the results, especially concerned with equivalence transformations, that will be used throughout this paper.

In the framework of Lie group analysis, given a system of differential equations, say
\begin{equation}
\Delta\left(\mathbf{x},\mathbf{u},\mathbf{u}^{(r)}\right)=0,
\label{sourcesystem}
\end{equation}
where $\mathbf{x}\in \mathbb{R}^n$ is the set of the independent variables,
$\mathbf{u}\in \mathbb{R}^m$ the set of the dependent variables,
and $\mathbf{u}^{(r)}$ the set of all partial derivatives of the $\mathbf{u}$'s with respect to
the $\mathbf{x}$'s up to the order $r$,
one is interested to find the admitted group of Lie symmetries.
Lie group analysis provides a powerful and unified approach to differential equations, both  ordinary and partial; for ordinary differential equations it algorithmically leads to lowering the order (or reducing to quadrature), for partial differential equations they allow for the construction of group--invariant solutions or
for introducing invertible point transformations mapping the differential equations in equivalent forms.

In many situations we have differential equations involving arbitrary elements (constants or functions), so that one has a class of differential equations. Here we shall consider a
class $\mathcal{E}(\mathbf{p})$ of first order PDEs involving some arbitrary continuously differentiable functions $p_k(\mathbf{x},\mathbf{u})$ ($k=1,\ldots,\ell$),
\begin{equation}
 \boldsymbol\Delta(\mathbf{x},\mathbf{u},\mathbf{u}^{(1)};\mathbf{p},\mathbf{p}^{(1)})=0,
\end{equation}
whose elements are given once we fix the functions $p_k$ ($\mathbf{p}^{(1)}$ denotes the set of first order partial derivatives of the $\mathbf{p}$'s with respect to their arguments).
To face this problem, it is convenient to consider equivalence transformations, \emph{i.e.},
transformations that preserve the differential structure of the equations in the
system but may change the form of the constitutive functions and/or parameters \cite{Ovsiannikov,IbragimovTorrisiValenti,Lisle,IbragimovTorrisi1994,TorrisiTracinaproceedings1996,TorrisiTracina1998,OzerSuhubi,OliveriSpeciale2012,OliveriSpeciale2013}.

\begin{definition}[Equivalence transformations \cite{Ovsiannikov}]
A one--parameter Lie group of equivalence transformations
of a family $\mathcal{E}(\mathbf{p})$ of PDEs is a one--parameter Lie group of transformations
given by
\begin{equation} 
\mathbf{X}= \mathbf{X}(\mathbf{x}, \mathbf{u},\mathbf{p}; a), \qquad
\mathbf{U}= \mathbf{U}(\mathbf{x}, \mathbf{u},\mathbf{p}; a), \qquad
\mathbf{P} =\mathbf{P}(\mathbf{x}, \mathbf{u},\mathbf{p}; a),
\end{equation}
$a$ being the parameter, which is locally a $C^\infty$ diffeomorphism and maps a class $\mathcal{E}(\mathbf{p})$ of differential equations
into itself; thus, it may change the differential equations (the form of the arbitrary elements therein involved) but preserves the differential structure.
\end{definition}
In the following we shall assume that the transformations of the independent and dependent variables do not involve the arbitrary elements $\mathbf{p}$.

In an augmented space  $\mathcal{A}\equiv\mathbb{R}^n  \times \mathbb{R}^m \times  \mathbb{R}^\ell$ \cite{Ovsiannikov,Lisle}, where the independent variables, the dependent variables and the arbitrary functions live, respectively,
the generator of the equivalence transformation,
\begin{equation}
\label{equivalenceoperator}
\Xi  =\sum_{i=1}^n\xi ^{i}(\mathbf{x}, \mathbf{u})\frac{\partial}{\partial x_{i}}
+ \sum_{A=1}^m\eta ^{A}(\mathbf{x}, \mathbf{u})\frac{\partial}{\partial u_A}
+\sum_{k=1}^{\ell}\mu^{k}(\mathbf{x}, \mathbf{u},\mathbf{p})\frac{\partial}{\partial p_k},
\end{equation}
involves also the infinitesimals $\mu^{k}(\mathbf{x}, \mathbf{u},\mathbf{p})$ accounting for the arbitrary functions  $p_k$. The search for continuous
equivalence transformations can be exploited by using the Lie infinitesimal criterion \cite{Ovsiannikov}.

The first prolongation of $\Xi$ writes as
\begin{equation}
\Xi^{(1)}= \Xi+
\sum_{A=1}^m\sum_{i=1}^n \eta^A_{[i]} \frac{\partial}{\partial u_{A,i}}+\sum_{k=1}^\ell \sum_{\alpha=1}^{n+m} \mu_{[\alpha]}^k\frac{\partial}{\partial p_{k,\alpha}},
\end{equation}
with
\begin{equation}
\eta_{[i]}^A= \frac{D\eta^A}{Dx_i}-\sum_{j=1}^n u_{A,j}\frac{D\xi^j}{Dx_i},\qquad
\mu_{[\alpha]}^k= \frac{\widetilde{D}\mu^k}{\widetilde{D}z_\alpha}-\sum_{\beta=1}^{n+m}p_{k,\beta}\frac{\widetilde{D}\zeta^\beta}{\widetilde{D}z_\alpha},
\end{equation}
($u_{A,j}=\frac{\partial u^A}{\partial x_j}$, $p_{k,\alpha}=\frac{\partial p_k}{\partial z_\alpha}$, $\mathbf{z}=(\mathbf{x},\mathbf{u})$, $\boldsymbol{\zeta}=(\boldsymbol{\xi},\boldsymbol{\eta})$),
where the \emph{Lie derivatives} are
\begin{equation}
\frac{D}{Dx_i}=\frac{\partial}{\partial x_i}+ \sum_{A=1}^m u_{A,i} \frac{\partial}{\partial u_A},\qquad
\frac{\widetilde{D}}{\widetilde{D}z_\alpha}=\frac{\partial}{\partial z_\alpha}+ \sum_{k=1}^\ell p_{k,\alpha} \frac{\partial}{\partial p_k}.
\end{equation}

\subsection{Finite transformations in the projected space}

In the augmented space $\mathcal{A}$, the arbitrary functions determining the class of differential equations
are assumed as dependent variables, and we require the invariance of the class in this augmented space.
If  we project the symmetries on the  space $\mathcal{Z}\equiv\mathbb{R}^n\times\mathbb{R}^m$ of the independent and dependent variables
(this is possible because the infinitesimals of independent and dependent variables are assumed to be independent of $\mathbf{p}$),
we obtain a transformation
changing an element of the class of differential equations to another element in the same class
(same differential structure but in general different arbitrary elements).
Such projected transformations map solutions
of a system in the class to solutions of a transformed system in the same class.

Thus, in the augmented space $\mathcal{A}$, given the equivalence generator \eqref{equivalenceoperator}
the integration of Lie's equations
\begin{equation}
\label{eq:Lie_equations_equivalence}
\begin{aligned}
&\frac{d\mathbf{X}}{da}={\boldsymbol\xi}(\mathbf{X},\mathbf{U}), \quad
\frac{d\mathbf{U}}{da}={\boldsymbol\eta}(\mathbf{X},\mathbf{U}),\quad
\frac{d\mathbf{P}}{d a}={\boldsymbol\mu}(\mathbf{X},\mathbf{U}, \mathbf{P}),\\
&\mathbf{X}(0)=\mathbf{x},\quad \mathbf{U}(0)=\mathbf{u},\quad\mathbf{P}(0)=\mathbf{p}
\end{aligned}
\end{equation}
provides the finite transformation which maps the class into itself.
On the contrary, the integration of the Lie's equations (\ref{eq:Lie_equations_equivalence}) in the projected space
$\mathcal{Z}$
gives an equivalence transformation
mapping a system in the class into another system in the same class.

\section{The model equations}
\label{model}
Consider the class $\mathcal{E}(\mathbf{p})$ with $\mathbf{p}= (p_1,\ldots,p_{16})$ 
of systems
\begin{equation}
\begin{aligned}
&\partial_{x_1} u_1+\partial_{x_2} u_2+\partial_{x_3} u_3+\partial_{x_4} u_4=0, \\
&\partial_{x_1} u_2+\partial_{x_2} p_1+\partial_{x_3} p_2+\partial_{x_4} p_3=p_{13},\\
&\partial_{x_1} u_3+\partial_{x_2} p_4+\partial_{x_3} p_5+\partial_{x_4} p_6=p_{14},\\
&\partial_{x_1} u_4+\partial_{x_2} p_7+\partial_{x_3} p_8+\partial_{x_4} p_9=p_{15},\\
&\partial_{x_1} u_5+\partial_{x_2} p_{10}+\partial_{x_3} p_{11}\partial_{x_4} p_{12}=p_{16},
\end{aligned}
\label{sis}
\end{equation}
where $\mathbf{x}\equiv(x_1,x_2,x_3,x_4)$ are the independent variables, 
$\mathbf{u}\equiv(u_1,u_2,u_3,u_4,u_5)$ the dependent variables, whereas 
$\mathbf{p}\equiv(p_1,\ldots, p_{16})$ stand for arbitrary continuously differentiable functions of $\mathbf{x}$ and $\mathbf{u}$. 
For instance, three-dimensional Euler equations of ideal gas--dynamics fall into the class (\ref{sis}). 

By requiring the invariance of the class $\mathcal{E}(\mathbf{p})$ in the augmented space
$\mathcal{A}\equiv\mathbb{R}^4\times\mathbb{R}^5\times\mathbb{R}^{16}$ through the Lie's infinitesimal criterion \cite{Ovsiannikov}, 
we determine 24 symmetry operators, whose expression is too long to be written here. In view of the results we want to achieve, we report the
projections of the admitted operators on the space 
$\mathcal{Z}\equiv\mathbb{R}^4\times\mathbb{R}^5$:
\begin{equation}
\label{operators}
\begin{aligned}
&\Xi_{1}=f_1(x_1)\partial_{x_1}-f_1^\prime(x_1) (u_2\partial_{u_2}+u_3\partial_{u_3}+u_4\partial_{u_4}),\\
&\Xi_{i}=f_i(\mathbf{x})\partial_{x_i}+\sum_{k=1}^4\left(u_k\partial_{x_k}f_i(\mathbf{x}) \partial_{u_i}-u_k\partial_{x_i}f_i(\mathbf{x}) \partial_{u_k}\right),\qquad (i=2,3,4),\\
&\Xi_{4+i}=u_if_{4+i}(\mathbf{x})\partial_{u_5}, \qquad (i=1,\ldots,5),\\
&\Xi_{10}=f_{10}(\mathbf{x})\partial_{u_5},\qquad
\Xi_{11}=\sum_{k=1}^4 f_{10+k}(\mathbf{x})\partial_{u_k},\qquad
\Xi_{12}=\sum_{k=1}^{4}u_k\partial_{u_k},
\end{aligned}
\end{equation}
where the functions $f_i$ ($i=1,\ldots,14$) are arbitrary functions of the indicated variables, along with the condition
\begin{equation}
\sum_{k=1}^4 \partial_{x_k} f_{10+k}(\mathbf{x})=0,
\end{equation}
and the prime ${}^\prime$ denotes the differentiation with respect to the argument.

By considering the corresponding Lie's equations we will be able to compute the finite corresponding transformations, say
\begin{equation}
\label{finite}
\mathbf{X}=\mathbf{X}(\mathbf{x},\mathbf{u};a), \qquad 
\mathbf{U}=\mathbf{U}(\mathbf{x},\mathbf{u};a)
\end{equation}
allowing us to map the original system \eqref{sis}
to a different system with the same differential structure; in particular,
we are interested to the case where the target system is an autonomous system of conservation laws:
\begin{equation}
\begin{aligned}
&\partial_{X_1} U_1+\partial_{X_2} U_2+\partial_{X_3} U_3+\partial_{X_4} U_4=0, \\
&\partial_{X_1} U_2+\partial_{X_2} P_1+\partial_{X_3} P_2+\partial_{X_4} P_3=0,\\
&\partial_{X_1} U_3+\partial_{X_2} P_4+\partial_{X_3} P_5+\partial_{X_4} P_6=0,\\
&\partial_{X_1} U_4+\partial_{X_2} P_7+\partial_{X_3} P_8+\partial_{X_4} P_9=0,\\
&\partial_{X_1} U_5+\partial_{X_2} P_{10}+\partial_{X_3} P_{11}+\partial_{X_4} P_{12}=0,\\
\end{aligned}
\label{sisbuono}
\end{equation}
where $P_i\equiv P_i(U_1,U_2,U_3,U_4,U_5)$, $i=1,\ldots,12$. Of course, a given system 
falling in the class~(\ref{sis}) can be mapped by an equivalence transformation to a system 
having the form~(\ref{sisbuono}) provided that the functions $p_i(\mathbf{x},\mathbf{u})$ 
($i=1,\ldots,16$) have special functional forms.
To simplify the computation, we exchange the source and target system; in fact,
taking the inverse transformation of \eqref{finite} (which is obtained by exchanging lower 
and capital letters and replacing $a$ with $-a$), and starting from the autonomous  system 
\eqref{sisbuono} of conservation laws, we are able to obtain the
equivalent nonautonomous system of balance laws. In such a way, we are able to identify, 
for a given equivalence transformation, the elements of the class~(\ref{sis}) that can be 
mapped to a system of autonomous conservation laws.

Now, since we start from an autonomous system of conservation laws to arrive to a 
nonautonomous system of balance laws, let us write the operators (\ref{operators})
in terms of the capital letters; then, we build the corresponding
finite transformations. 

The most general finite transformation
can be recovered by composition of the finite transformations induced by each generator.

\subsection{Finite transformations generated by $\Xi_1$}
By considering the generator  $\Xi_1$,
\begin{equation}
\Xi_{1}=f(X_1)\partial_{X_1}-f^\prime(X_1)\left(U_2\partial_{U_2}+U_3\partial_{U_3}+U_4\partial_{U_4}\right),
\end{equation}
where we set $f(X_1)=f_1(X_1)$, we get the finite transformation
\begin{equation}
\begin{aligned}
&x_1=\tilde{x}_1(X_1;a),\qquad x_2=X_2,\qquad
x_3=X_3,\qquad x_4=X_4,\\
&u_1=U_1,\qquad u_2=U_2\frac{f(X_1)}{f(x_1)},\qquad
u_3=U_3\frac{f(X_1)}{f(x_1)},\qquad
u_4=U_4\frac{f(X_1)}{f(x_1)},\qquad
u_5=U_5,
\end{aligned}
\end{equation}
$\tilde{x}_1(X_1;a)$ being such that 
$\displaystyle\partial_{X_1}\tilde{x}_1=\frac{f(x_1)}{f(X_1)}$,
whereupon we may write
\begin{equation}
U_1=u_1,\qquad
U_2=u_2\partial_{X_1}\tilde{x}_1,\qquad
U_3=u_3\partial_{X_1}\tilde{x}_1,\qquad
U_4=u_4\partial_{X_1}\tilde{x}_1,\qquad
U_5=u_5,
\end{equation}
and system (\ref{sis}) is equivalent to (\ref{sisbuono}) with
\begin{equation}
\begin{aligned}
&p_k=\frac{P_k}{(\partial_{X_1}\tilde{x}_1)^{2}},\qquad k=1,\ldots,12,\\
&p_{13}=-u_2\frac{\partial^2_{X_1X_1}\tilde{x}_1}{(\partial_{X_1}\tilde{x}_1)^{2}},\quad p_{14}=-u_3\frac{\partial^2_{X_1X_1}\tilde{x}_1}{(\partial_{X_1}\tilde{x}_1)^{2}},\quad p_{15}=-u_4\frac{\partial^2_{X_1X_1}\tilde{x}_1}{(\partial_{X_1}\tilde{x}_1)^{2}},\quad p_{16}=0,
\end{aligned}
\end{equation}
where $P_k=P_k(u_1,u_2\partial_{X_1}\tilde{x}_1,u_3\partial_{X_1}\tilde{x}_1,u_4\partial_{X_1}\tilde{x}_1,u_5)$, $k=1,\ldots,12$.

\subsection{Finite transformations generated by $\Xi_2$, $\Xi_3$ and $\Xi_4$}
By taking the generators $\Xi_i$, $i=2,3,4$,
\begin{equation}
\Xi_{i}=f(\mathbf{X})\partial_{X_i}+\sum_{k=1}^4\left(U_k\partial_{X_k}f(\mathbf{X})\partial_{U_i}-U_k\partial_{X_i}f(\mathbf{X})\partial_{U_k}\right),\qquad (i=2,3,4),
\end{equation}
where we set $f(\mathbf{X})=f_i(\mathbf{X})$,
we may write the general finite transformation arising from the integration of Lie's equations in the three cases in a unified form:
\begin{equation}
\begin{aligned}
&x_k=\left\{
  \begin{aligned}
  & X_k, \quad && k=1,\ldots,4,\; k\neq i \\
  &\tilde{x}_k(\mathbf{X};a),  \quad &&  k= i,
  \end{aligned}
  \right.\\
&u_k=\left\{
  \begin{aligned}
  &U_k\frac{f(\mathbf{X})}{f(\mathbf{x})},  \quad && k=1,\ldots,4,\; k\neq i,\\
  &U_k+f(\mathbf{X})\sum_{j=1,j\neq i}^4 U_j \int_{0}^{a}\frac{\partial_{X_j}f(\mathbf{x})}{f(\mathbf{x})}da, \quad &&    k=i,\\
  &U_k,  \quad && k=5,  
  \end{aligned}
  \right.
\end{aligned}
\end{equation}
where $\tilde{x}_i(\mathbf{X};a)$  is such that
\begin{equation}
\partial_{X_k}\tilde{x}_i=\left\{
\begin{aligned}
&f(\mathbf{x})\int_0^a \frac{\partial_{X_k}f(\mathbf{x})}{f(\mathbf{x})}da,  \quad && k\neq i\\
&\frac{f(\mathbf{x})}{f(\mathbf{X})},  \quad && k=i.
\end{aligned}
\right.
\end{equation}
By introducing the matrix $J$ with the $(j,k)$--entry equal to $\partial_{X_k}\tilde{x}_j$ ($j,k=1,\ldots,4$), the $(5,5)$--entry equal to $\partial_{X_i}\tilde{x}_i$ and all remaining entries vanishing, we may write
\begin{equation}
\label{Uk2}
\mathbf{u}=A\mathbf{U}, \qquad A=\frac{J}{\partial_{X_i}\tilde{x}_i};
\end{equation}
moreover, by defining the matrices
\[
q=\left[
\begin{array}{ccccc}
u_1 & u_2 & u_3 & u_4 & 0\\
u_2 & p_1 & p_2 & p_3 & 0\\
u_3 & p_4 & p_5 & p_6 & 0\\
u_4 & p_7 & p_8 & p_9 & 0\\
u_5 & p_{10} & p_{11} & p_{12} & 0
\end{array}
\right],
\qquad
Q=\left[
\begin{array}{ccccc}
U_1 & U_2 & U_3 & U_4 & 0\\
U_2 & P_1 & P_2 & P_3 & 0\\
U_3 & P_4 & P_5 & P_6 & 0\\
U_4 & P_7 & P_8 & P_9 & 0\\
U_5 & P_{10} & P_{11} & P_{12} & 0
\end{array}
\right], 
\]
system \eqref{sisbuono} is mapped to system \eqref{sis} with 
\begin{equation}
\begin{aligned}
&q=AQJ^T,\qquad \\&p_{11+m}=\sum_{j=1}^5A_{mj}\sum_{\ell=1}^5\left(\sum_{k=1}^4 u_k \frac{\partial^2R_{\ell j}}{\partial U_\ell \partial X_k}-\sum_{k=1}^5 u_k \frac{\partial^2R_{\ell j}}{\partial U_k\partial X_\ell}\right), \quad m=2,\ldots,5,
\end{aligned}
\label{qQ}
\end{equation}
where $R_{\ell j}$ is the generic entry of the matrix $JP^T$, and it is $P_k=P_k(\mathbf{U})$, ($k=1,\dots 12$), with $\mathbf{U}$ defined by (\ref{Uk2}); note that the right hand side
of (\ref{qQ})$_2$ is vanishing for $m=1$.

\subsection{Finite transformations generated by $\Xi_5$, $\Xi_6$, $\Xi_7$, $\Xi_8$}
By considering the generator  $\Xi_{4+i}$ ($i=1,\ldots,4$),
\begin{equation}
\Xi_{4+i}=  U_if(\mathbf{X})\partial_{U_5},
\end{equation}
where we set $f(\mathbf{X})=f_{4+i}(\mathbf{X})$,
we get from Lie's equations the finite transformation
\begin{equation}
x_k=X_k, \qquad u_k=U_K, \qquad k=1,\ldots,4 ,\qquad
u_5=U_5 -a U_i f(\mathbf{X}).
\end{equation}
System (\ref{sisbuono}) is equivalent to system (\ref{sis}) if
\begin{equation}
\begin{aligned}
&p_k=P_k, \qquad k=1,\ldots 9,\\
&p_{9+k}=\left\{
\begin{aligned}
&P_{9+k}+au_{k+1} f(\mathbf{x}),\quad &&i=1,\\
&P_{9+k}+aP_{3i+k-6} f(\mathbf{x}),\quad &&i=2,3,4,
\end{aligned}
\right.\qquad k=1,2,3,\\
&p_{12+k}=au_i\sum_{j=2}^4\partial_{X_j}f(\mathbf{x})\partial_{U_5}P_{3k+j-4}\qquad k=1,2,3,\\
&p_{16}=\left\{
\begin{aligned}
&au_i\left(\partial_{X_1}f(\mathbf{x})+\sum_{j=2}^4\partial_{X_j}f(\mathbf{x})\partial_{U_5}P_{8+j} \right),\quad && i=1\\
&au_i\left(\partial_{X_1}f(\mathbf{x})+\sum_{j=2}^4\partial_{X_j}f(\mathbf{x})\left(\partial_{U_5}P_{8+j}
+af(\mathbf{x})\partial_{U_5}P_{3i+j-7}\right)\right),\quad && i=2,3,4,\\
\end{aligned}
\right.
\end{aligned} 
\end{equation}
where $P_k=P_k(u_1, u_2, u_3,u_4,u_5+au_i f(\mathbf{x}))$, $k=1,\ldots,12$.

\subsection{Finite transformations generated by $\Xi_9$}
By considering the generator  $\Xi_{9}$,
\begin{equation}
\Xi_{9}=  U_5f(\mathbf{X}) \partial_{U_5},
\end{equation}
where we set $f(\mathbf{X})=f_{9}(\mathbf{X})$,
we get from Lie's equations the finite transformation
\begin{equation}
 x_k=X_k,\qquad u_k=U_k,\qquad k=1,\ldots,4,\qquad u_5=U_5\exp(af(\mathbf{X})).
\end{equation}
The system (\ref{sisbuono}) is equivalent to system (\ref{sis}) provided that:
\begin{equation}
\begin{aligned}
&p_k=P_k,\qquad k=1,\ldots,12,\\
&p_{12+k}=a\exp{(-af(\mathbf{x}))}\left(\sum_{j=2}^4\partial_{X_j}f(\mathbf{x})\partial_{U_5}P_{3k+j-4}\right)u_5,\qquad k=1,2,3,\\
&p_{16}=a\left(\partial_{X_1}f(\mathbf{x})+\sum_{j=2}^4\partial_{X_j}f(\mathbf{x})\partial_{U_5}P_{8+j}\right)u_5,
\end{aligned}
\end{equation}
where $P_k=P_k(u_1,u_2,u_3,u_4,u_5\exp(-af(\mathbf{x})))$, $i=1,\ldots,12$.

\subsection{Finite transformations generated by $\Xi_{10}$}
By considering the generator  $\Xi_{10}$,
\begin{equation}
\Xi_{10}=  f(\mathbf{X})\partial_{U_5},
\end{equation}
where we set $f(\mathbf{X})=f_{10}(\mathbf{X})$,
we get from Lie's equations the finite transformation
\begin{equation}
x_k=X_k,\qquad u_k=U_k,\qquad k=1,\ldots,4,\qquad
u_5=U_5+af(\mathbf{X}),
\end{equation}
and the equivalence between (\ref{sis}) and (\ref{sisbuono}) is recovered provided that
\begin{equation}
\begin{aligned}
&p_k=P_k,\qquad k=1,\ldots,12,\\
&p_{12+k}=a\sum_{j=2}^4\partial_{X_j}f(\mathbf{x})\partial_{U_5}P_{3k+j-4},\qquad k=1,2,3,\\
&p_{16}=a\left(\partial_{X_1}f(\mathbf{x})+\sum_{j=2}^4\partial_{X_j}f(\mathbf{x})\partial_{U_5}P_{8+j}\right),
\end{aligned}
\end{equation}
where $P_k=P_k(u_1,u_2,u_3,u_4,u_5-af(\mathbf{x}))$, $k=1,\ldots,12$.

\subsection{Finite transformations generated by $\Xi_{11}$}
By considering the generator  $\Xi_{11}$,
\begin{equation}
\sum_{k=1}^4 g_{k}(\mathbf{X})\partial_{U_k},
\end{equation}
where we set $g_k(\mathbf{X})=f_{10+k}(\mathbf{X})$, along with the constraint
$\partial_{X_k}g_k(\mathbf{x})=0$,
and integrating the Lie's equations, the following finite transformation arises:
\begin{equation}
x_k=X_k,\qquad u_k=U_k+ag_k(\mathbf{x}), \qquad k=1,\ldots,4,\qquad u_5=U_5.
\end{equation}
System \eqref{sisbuono} is equivalent to system \eqref{sis} provided that:
\begin{equation}
\begin{aligned}
&p_k=P_k,\qquad &&k=1,\ldots,12,\\
&p_{12+k}=a\left(\partial_{X_1}g_{k+1}(\mathbf{x})+\sum_{i=1}^4\sum_{j=2}^4\partial_{X_j}g_i(\mathbf{x})\partial_{U_i}P_{3k+j-4}\right),\qquad &&k=1,2,3,\\
&p_{16}=a\left(\sum_{i=1}^4\sum_{j=2}^4\partial_{X_j}g_i(\mathbf{x})\partial_{U_i}P_{8+j}\right),\\
\end{aligned}
\end{equation}
where $P_k=P_k(u_1-ag_1(\mathbf{x}),u_2-ag_2(\mathbf{x}),u_3-ag_3(\mathbf{x}),u_4-ag_4(\mathbf{x}),u_5)$.

\subsection{Equivalence transformations generated by $\Xi_{12}$}
In this case the finite transformation consists of a uniform scaling of the dependent variables,
\begin{equation}
\begin{aligned}
&\mathbf{x}=\mathbf{X}, \qquad \mathbf{u}=\exp(a)\mathbf{U},
\end{aligned}
\end{equation}
and for such a transformation there are no balance laws equivalent to conservation laws.

\section{Physical application}
\label{applications}
In this Section,  we make some assumptions on the form of the functions involved in the generators of equivalence transformations
in order to deal with physically relevant systems of differential equations. In particular, we construct the finite transformations corresponding to  the
infinitesimal generator $\sum_{i=1}^4\Xi_i$,  where we assume
\[
\begin{aligned}
 f_2=n_1(X_1) X_2+n_2(X_1) X_3, \quad f_3=-n_2(X_1)X_2 +n_1(X_1)X_3,\quad f_4=n_3(X_1), 
\end{aligned}
\] 
with $n_i(X_1)$, $i=1,\ldots,3$, arbitrary functions of $X_1$.

Integration of Lie's equations provides:
\begin{equation}
\label{generalforeuler}
\begin{aligned}
&x_1=\tilde{x}_1(X_1;a),\qquad x_4=\tilde{x}_4(X_1,X_4;a)=X_4+m_3(X_1;a),\\
&x_2=\tilde{x}_2(X_1,X_2,X_3;a)=\exp(m_1(X_1;a))(X_2\cos(m_2(X_1;a))+ X_3 \sin(m_2(X_1;a))),\\
&x_3=\tilde{x}_3(X_1,X_2,X_3;a)=\exp(m_1(X_1;a))(-X_2\sin(m_2(X_1;a))+ X_3 \cos(m_2(X_1;a))),\\
&U_1=\exp(2 m_1(X_1;a))u_1,\\
&U_2=\exp(m_1(X_1;a))\left[(u_2 \cos(m_2(X_1;a))-u_3\sin(m_2(X_1;a)))\partial_{X_1}\tilde{x}_1\right.\\
&\qquad \left.-u_1(\partial_{X_1}\tilde{x}_2\cos(m_2(X_1;a))-\partial_{X_1}\tilde{x}_3\sin(m_2(X_1;a)))\right],\\
&U_3=\exp(m_1(X_1;a))\left[(u_2 \sin(m_2(X_1;a))+u_3\cos(m_2(X_1;a)))\partial_{X_1}\tilde{x}_1\right.\\
&\qquad\left.-u_1(\partial_{X_1}\tilde{x}_2\sin(m_2(X_1;a))+\partial_{X_1}\tilde{x}_3\cos(m_2(X_1;a)))\right],\\
&U_4=\exp(2 m_1(X_1;a))\left(u_4\partial_{X_1}\tilde{x}_1-u_1\partial_{X_1}\tilde{x}_4\right),\qquad
U_5=u_5,
\end{aligned}
\end{equation}
where
\[
\begin{aligned}
&m_i(X_1;a)=\int^{x_1}_{X_1}\frac{n_i(s)}{f_1(s)}ds, \qquad &&\tilde{n}_i(X_1;a)=n_i(x_1)-n_i(X_1),\qquad i=1,2,3,\\
&\partial_{X_1}\tilde{x}_1=\frac{f_1(x_1)}{f_1(X_1)}, \qquad 
&& \partial_{X_1}\tilde{x}_2=\frac{ \tilde{n}_1(X_1;a)x_2+\tilde{n}_2(X_1;a)x_3}{f_1(X_1)},\\
&\partial_{X_1}\tilde{x}_3=\frac{ -\tilde{n}_2(X_1;a)x_2+\tilde{n}_1(X_1;a)x_3}{f_1(X_1)}, \qquad
&&\partial_{X_1}\tilde{x}_4=\frac{ \tilde{n}_3(X_1;a)}{f_1(X_1)}.
\end{aligned}
\]
System (\ref{sisbuono}) describes the 3D unsteady flow of an ideal fluid subject to no extraneous force along with the choices
\begin{equation*}
\begin{aligned}
&U_1=\rho, \quad U_2=\rho u,\quad U_3=\rho v, \quad U_4=\rho w, 
\qquad U_5=\rho S,\\
&P_1=\frac{U_{2}^2}{U_1}+p(U_1,U_5),\quad P_2=P_4=\frac{U_{2}U_{3}}{U_1},\quad P_3=P_7=\frac{U_{2}U_{4}}{U_1},\\
& P_5=\frac{U_{3}^2}{U_1}+p(U_1,U_5),\quad P_6=P_8=\frac{U_{3}U_{4}}{U_1},\quad P_9=\frac{U_{4}^2}{U_1}+p(U_1,U_5),\\
&P_{10}=\frac{U_{2}U_{5}}{U_1},\quad P_{11}=\frac{U_{3}U_{5}}{U_1},\quad P_{12}=\frac{U_{4}U_{5}}{U_1},
\end{aligned}
\end{equation*}
$\rho$ being the fluid mass density, $(u,v,w)$ the components of its velocity, $S$ the entropy, and $p(\rho,S)$ the pressure. 
Thorugh the transformation (\ref{generalforeuler}) we get the system (\ref{sis}) with
\begin{equation*}
\begin{aligned}
&p_1=\frac{u_{2}^2}{u_1}+\frac{p(\exp(-2 m_1)u_1,u_5)}{(\partial_{X_1}\tilde{x}_1)^2},\quad 
p_2=p_4=\frac{u_{2}u_{3}}{u_1},\quad p_3=p_7=\frac{u_{2}u_{4}}{u_1},\\ 
&p_5=\frac{u_{3}^2}{u_1}+\frac{p(\exp(-2 m_1)u_1,u_5)}{(\partial_{X_1}\tilde{x}_1)^2},\quad p_6=p_8=\frac{u_{3}u_{4}}{u_1},\quad
p_9=\frac{u_{4}^2}{u_1}+\frac{p(\exp(-2 m_1)u_1,u_5)}{(\partial_{X_1}\tilde{x}_1)^2},\\
&p_{10}=\frac{u_{2}u_{5}}{u_1},\quad p_{11}=\frac{u_{3}u_{5}}{u_1},\quad p_{12}=\frac{u_{4}u_{5}}{u_1},\\
&p_{13}=2\frac{\partial_{X_1}m_{1}u_2+\partial_{X_1}m_{2}u_3}{\partial_{X_1}\tilde{x}_1}-\frac{\partial^2_{X_1X_1}\tilde{x}_1}{(\partial_{X_1}\tilde{x}_1)^2}u_2\\
&\qquad+\frac{x_2(\partial^2_{X_1X_1}m_{1}-(\partial_{X_1}m_{1})^2+(\partial_{X_1}m_{2})^2)+x_3(\partial^2_{X_1X_1}m_{2}-2\partial_{X_1}m_{1}\partial_{X_1}m_{2})}{(\partial_{X_1}\tilde{x}_1)^2}u_1,\\
&p_{14}=2\frac{\partial_{X_1}m_{1}u_3-\partial_{X_1}m_{2}u_2}{\partial_{X_1}\tilde{x}_1}-\frac{\partial^2_{X_1X_1}\tilde{x}_1}{(\partial_{X_1}\tilde{x}_1)^2}u_3\\
&\qquad+\frac{-x_2(\partial^2_{X_1X_1}m_{2}-2\partial_{X_1}m_{1}\partial_{X_1}m_{2})+x_3(\partial^2_{X_1X_1}m_{1}-(\partial_{X_1}m_{1})^2+(\partial_{X_1}m_{2})^2)}{(\partial_{X_1}\tilde{x}_1)^2}u_1,\\
&p_{15}=-\frac{\partial^2_{X_1X_1}\tilde{x}_1}{(\partial_{X_1}\tilde{x}_1)^2}u_4+\frac{\partial^2_{X_1X_1}\tilde{x}_4}{(\partial_{X_1}\tilde{x}_1)^2}u_1,\qquad p_{16}=2\frac{\partial_{X_1}m_{1}}{\partial_{X_1}\tilde{x}_1}u_5.
\end{aligned}
\end{equation*}

By choosing $\partial_{X_1}\tilde{x}_1=1$ (whereupon $x_1=X_1+a$), $m_1=0$, 
$m_2=\omega X_1+X_{1_0}$, $m_3=\frac{gX_1^2}{2}+a_1X_1+a_0$,
where $\omega$, $a_0$, $a_1$, $a_2$, $g$ and $X_{1_0}$ are constants, we recover
the system
\begin{equation}
\begin{aligned}
&\partial_{x_1} u_1+\partial_{x_2} u_2+\partial_{x_3} u_3+\partial_{x_4} u_4=0, \\
&\partial_{x_1} u_2+\partial_{x_2}\left(\frac{u_{2}^2}{u_1}+p(u_1,u_5)\right)
+\partial_{x_3}\left(\frac{u_{2}u_{3}}{u_1}\right)+\partial_{x_4}\left(\frac{u_{2}u_{4}}{u_1}\right)=2\omega u_3-\omega^2 x_2 u_1,\\
&\partial_{x_1}u_3+\partial_{x_2}\left(\frac{u_{2}u_{3}}{u_1}\right)
+\partial_{x_3}\left(\frac{u_{3}^2}{u_1}+p(u_1,u_5)\right)
+\partial_{x_4}\left(\frac{u_{3}u_{4}}{u_1}\right)=-2\omega u_2+\omega^2 x_3 u_1,\\
&\partial_{x_1}u_4+\partial_{x_2}\left(\frac{u_{2}u_{4}}{u_1}\right)
+\partial_{x_3}\left(\frac{u_{3}u_{4}}{u_1}\right)
+\partial_{x_4}\left(\frac{u_{4}^2}{u_1}+p(u_1,u_5)\right)=g u_1,\\
&\partial_{x_1}u_5+\partial_{x_2}\left(\frac{u_{2}u_{5}}{u_1}\right)+
\partial_{x_3}\left(\frac{u_{3}u_{5}}{u_1}\right)+
\partial_{x_4}\left(\frac{u_{4}u_{5}}{u_1}\right)=0.
\end{aligned}
\end{equation}
With obvious identifications, we recognize the equations of an ideal gas in a
non--inertial frame rotating with constant angular velocity $\omega$ around the vertical $x_4$--axis and subject to gravity.
This implies that the Euler equations for an ideal gas in a non--inertial frame rotating with constant angular velocity around a vertical axis and subject to gravity can be transformed in a form where the gravity and apparent forces disappear.

\section{Conclusions}
\label{conclusions}
In this paper we have characterized classes of PDEs in four independent variables expressed under the form of a linear conservation law and four
nonautonomous nonlinear balance laws that can be transformed by an invertible point transformation into an autonomous system of conservation laws.
This has been accomplished through the use of
equivalence transformations. A physical application has been provided: it has been shown the equivalence of the 3D unsteady Euler equations of an ideal gas subject to gravity and Coriolis forces with the corresponding system where forces are absent.


\begin{thebibliography}{99}

\bibitem{Dafermos} C.~M.~Dafermos. Hyperbolic conservation laws in continuum physics.
Springer--Verlag, Berlin, 2000.

\bibitem{Bressan}Bressan, A., \emph{Hyperbolic systems of conservation laws}
(Oxford University Press, Oxford, 2000).

\bibitem{LeVeque} LeVeque, R.~J., \emph{Numerical Methods for Conservation Laws}
(Birkh\"auser, Basel, 1992).

\bibitem{Ovsiannikov}  Ovsiannikov, L.~V., \emph{Group analysis of differential equations} (Academic Press, New York, 1982).

\bibitem{Ibragimov} Ibragimov, N.~H.,  \emph{Transformation groups applied to mathematical physics}
(D. Reidel Publishing Company, Dordrecht, 1985).

\bibitem{Olver}  Olver, P.~J.,  \emph{Applications of Lie groups to differential equations} (Springer, New York, 1986).

\bibitem{Ibragimov:CRC} Ibragimov, N.~H., \emph{CRC Handbook of Lie group analysis of differential equations: Symmetries, Exact Solutions and Conservation Laws}
(CRC Press, Boca Raton, 1994, 1995, 1996).

\bibitem{Olver2}  Olver, P.~J., \emph{Equivalence, invariants, and symmetry}
(Cambridge University Press, New York, 1995).

\bibitem{Baumann} Baumann, G., \emph{Symmetry analysis of differential equations
with Mathematica} (Springer, New York, 2000).

\bibitem{Meleshko2005} Meleshko, S.~V., \emph{Methods for constructing exact solutions
of partial differential equations} (Springer, New York, 2005).

\bibitem{BlumanCheviakovAnco} Bluman, G.~W., Cheviakov, A.~F., Anco, S.~C.,
\emph{Applications of symmetry methods to partial differential equations}
(Springer, New York, 2009).

\bibitem{OliveriSpeciale1998} Oliveri, F., Speciale, M.~P., ``Exact solutions to the equations of ideal gas-dynamics by means of the substitution principle,'' Int. J. Non--linear Mech. \textbf{33},  585--592 (1998).

\bibitem{OliveriSpeciale1999} Oliveri, F., Speciale, M.~P., ``Exact solutions to the equations of perfect gases through Lie group analysis and substitution principles,'' Int. J. Non--linear Mech. \textbf{34}, 1077--1087 (1999).

\bibitem{OliveriSpeciale2002} Oliveri, F., Speciale, M.~P., ``Exact solutions to the unsteady equations of perfect gases through Lie group analysis and substitution principles,'' Int. J. Non--linear Mech. \textbf{37}, 257--274 (2002).

\bibitem{OliveriSpeciale2005} Oliveri, F., Speciale, M.~P., ``Exact solutions to the ideal magneto-gas-dynamics equations through Lie group analysis and substitution principles,'' J. Phys. A, Math. Gen. \textbf{38}, 8803--8820 (2005).

\bibitem{MargheritiSpeciale} Margheriti, L., Speciale, M.~P., ``Unsteady solutions of Euler equations generated by steady solutions,'' Acta Appl. Math. \textbf{113}, 289--303 (2011).

\bibitem{KumeiBluman} Kumei, S., Bluman, G.~W.,  ``When nonlinear differential equations are equivalent to linear differential equations,'' SIAM J. Appl. Math. \textbf{42}, 1157--1173 (1982).

\bibitem{DonatoOliveri1993} Donato, A., Oliveri, F., ``Reduction to autonomous form by group analysis and exact solutions of axi--symmetric MHD equations,'' Math. Computer Model. \textbf{18}, 83--90 (1993). 

\bibitem{DonatoOliveri1994}  Donato, A., Oliveri, F., ``Linearization procedure of nonlinear
first order systems of PDE's by means of canonical variables related to
Lie groups of point transformations,'' J. Math. Anal. Appl. \textbf{188} 552--568 (1994).

\bibitem{DonatoOliveri1995} Donato, A., Oliveri, F., ``When nonautonomous equations
are equivalent to autonomous ones,'' Appl. Anal. \textbf{58}, 313--323 (1995).

\bibitem{DonatoOliveri1996} Donato, A., Oliveri, F., ``How to build up variable
transformations allowing one to map nonlinear hyperbolic equations
into autonomous or linear ones,'' Transp. Th. Stat. Phys. \textbf{25}, 303--322 (1996).

\bibitem{CurroOliveri2008} Curr\`o, C., F.~Oliveri, F., ``Reduction of nonhomogeneous
quasilinear $2\times  2$ systems to homogeneous and autonomous form,''
J. Math. Phys. \textbf{49}, 103504-1--103504-11 (2008).

\bibitem{Oliveri2010} Oliveri, F., ``Lie symmetries of differential equations: classical results and recent contributions,'' Symmetry \textbf{2}, 658--706 (2010).

\bibitem{Oliveri2012} Oliveri, F., ``General dynamical systems described by first order quasilinear PDEs reducible to homogeneous and autonomous form.''
Int. J. Non-linear Mech. \textbf{47}, 53--60 (2012).

\bibitem{IbragimovTorrisiValenti} Ibragimov, N.~H., Torrisi, M., Valenti, A.,
``Preliminary group classification of equations $v_{tt} = f(x,v_{x})v_{xx}+g(x, v_x)$,'' J. Math. Phys. \textbf{32}, 2988--2995 (1991).

\bibitem{Lisle} Lisle, I.~G.,
\emph{Equivalence transformations for classes of differential  equations} (PhD dissertation, University of British Columbia, Vancouver B.C., Canada, 1992).

\bibitem{IbragimovTorrisi1994} Ibragimov, N.~H., Torrisi, M., ``Equivalence groups for balance equations,'' J. Math. Anal. Appl. \textbf{184}, 441--452 (1994).

\bibitem{TorrisiTracinaproceedings1996}
Torrisi, M., Tracin\`{a}, R. ``Equivalence transformations for system of first order quasilinear partial differential equations,'' In \emph{Modern Group Analysis VI, Developments in Theory, Computations and Applications} (Edited by N.~H.~Ibragimov and F.~H.~Mahomed), 115--135 (New Age International, New Delhi, 1997).

\bibitem{Meleshko1996}  Meleshko, S.~V., ``Generalization of the equivalence transformations'', J. Nonlinear Math. Phys.\textbf{3}, 170--174 (1996).

\bibitem{TorrisiTracina1998} Torrisi, M., Tracin\`{a}, R. ``Equivalence transformations and symmetries for a heat conduction model,'' Int. J. Non-linear Mech. \textbf{33}, 473--486 (1998).

\bibitem{OzerSuhubi} \"Ozer, S., Suhubi, E., ``Equivalence groups for first--order balance equations and applications to electromagnetism,'' Theor. Math. Phys. \textbf{137}, 1590--1597 (2003).

\bibitem{OliveriSpeciale2012} Oliveri, F., Speciale, M.~P., ``Equivalence transformations of quasilinear first order systems and reduction to autonomous and homogeneous form,''
Acta Appl. Math. \textbf{122}, 447--460 (2012).

\bibitem{OliveriSpeciale2013} Oliveri, F., Speciale, M.~P., ``Reduction of balance laws to conservation laws by means of equivalence transformations,'' J. Math. Phys. \textbf{54}, 041506 (2013).
\end{thebibliography}
\end{document}